# EPSAPG: A Pipeline Combining MMseqs2 and PSI-BLAST to Quickly Generate Extensive Protein Sequence Alignment Profiles

Issar Arab
Department of Computer Science, University of Antwerp, 2020 Antwerp, Belgium
Biomedical Informatics Network Antwerpen (biomina), 2020 Antwerp, Belgium
Email: issar.arab@tum.de

*Abstract*— Numerous machine learning (ML) models employed in protein function and structure prediction depend on evolutionary information, which is captured through multiple-sequence alignments (MSA) or position-specific scoring matrices (PSSM) as generated by PSI-BLAST. Consequently, these predictive methods are burdened by substantial computational demands and prolonged computing time requirements. The principal challenge stems from the necessity imposed on the PSI-BLAST software to load large sequence databases sequentially in batches and then search for sequence alignments akin to a given query sequence. In the case of batch queries, the runtime scales even linearly. The predicament at hand is becoming more challenging as the size of bio-sequence data repositories experiences exponential growth over time and as a consequence, this upward trend exerts a proportional strain on the runtime of PSI-BLAST. To address this issue, an eminent resolution lies in leveraging the MMseqs2 method, capable of expediting the search process by a magnitude of 100. However, MMseqs2 cannot be directly employed to generate the final output in the desired format of PSI-BLAST alignments and PSSM profiles. In this research work, I developed a comprehensive pipeline that synergistically integrates both MMseqs2 and PSI-BLAST, resulting in the creation of a robust, optimized, and highly efficient hybrid alignment pipeline. Notably, the hybrid tool exhibits a significant speed improvement, surpassing the runtime performance of PSI-BLAST in generating sequence alignment profiles by a factor of two orders of magnitude. It is implemented in C++ and is freely available under the MIT license at https://github.com/issararab/EPSAPG.

## I. INTRODUCTION

In computational biology, multiple sequence alignment methods, or MSA-methods, are a set of algorithmic solutions for the alignment of evolutionarily related sequences. They can be applied to DNA, RNA, or protein sequences. These algorithms are designed to incorporate evolutionary events such as mutations, insertions, deletions, and rearrangements under certain conditions [1]. Due to the complexity of the development of such a vital modeling tool which necessitates the addressing of both complex computational and biological problems, MSA has always been recognized as an NP-complete problem. That is the reason behind the numerous alternative algorithms built, more than 100 variants, aiming at providing an accurate MSA over the last four decades [2]. Despite the various alternatives of MSA methods, they all share a core asset: their reliance on estimated and typically greedy heuristics, enforced by the NP-complete nature of the problem. These heuristics are, more or less, dependent on specific data properties, like the length, the relatedness, the type of homology, and others.

The aim of MSA methods is to align any set of biological sequences, either RNA, DNA or proteins, in such a way they will capture their evolutionary, functional, or structural relationships. This is accomplished by adding gaps of different lengths within the sequences, enabling the homologous regions to be aligned with one another, which is analogous to aligning beads of similar colour in an Abacus. This is an interesting analogy simulated by the ancient counting frame, the Abacus, which was used as a calculator in Europe, China and Russia, centuries prior to the adoption of the written Arabic numerical scheme. From an evolutionary point of view, these gaps reflect insertions and deletions within the genome that are hypothesized or believed to have occurred during the evolution of sequences from a common ancestor [1].

One of the most famous and widely used MSA method in computational biology is PSI-BLAST [3, 4], which stands for the Position-Specific Iterative Basic Local Alignment Search Tool. It is a software that runs a multiple sequence alignment algorithm powered by the dynamic programming optimization paradigm to search a given database of protein sequences [5]. Given a query sequence, the algorithm retrieves homologous sequences that pass a predefined threshold. This threshold-based computational similarity method, which uses protein-protein BLAST [4, 5, 6] to identify regions of local alignment, is used to construct the Position-Specific Scoring Matrix (PSSM), a.k.a the sequence alignment profile, from the retrieved family of sequences. The PSSM profiles contain statistical representations of residues in a given sequence of a protein with respect to all its relevant aligned protein sequences (i.e. homologs) in the database. A profile captures the conservation pattern in an alignment and stores it as a matrix of scores for each position in the alignment [4], where high scores are assigned to highly conserved positions and lower scores to low conserved ones. Hence, the matrix representation includes what is known in the research community by protein evolutionary information.

In modern computational biology, particularly within the field of proteins and predicting their properties, sequence alignment and PSSM profiles represent the de facto standard input for almost all machine learning methods [7]. The main functionality of these alignment techniques is to search for homologs of a query sequence in a database of protein sequences, as they tend to share structure and function. For the past two decades, training machine learning models with evolutionary information representations, generated by multiple sequence alignments, has revolutionized the prediction power of AI methods. Multiple aspects of protein function and structure were studied and investigated following the same approach and achieved significant results in the prediction performance. Such downstream-specific tasks include protein secondary structure [8, 9, 10, 11], transmembrane protein regions [12, 13, 14], inter-residue contacts [15], and sub-cellular localization predictions [16, 17] as well as protein-to-protein interactions [18, 19, 20]. However, this increase in performance has become costly in recent years, with the continuous exponential growth of bio-sequence data pools. UniProt is one example of such datastores, in which the entries keep doubling every couple of years [21].

At present, MMseqs2 [22, 23], known as Many-against-Many Sequence Searching, emerges as a notable alternative MSA solution to handle such extensive data sets. However, its effective usage demands considerable hardware resources. Nevertheless, it's important to note that this solution doesn't generate sequence alignment profiles in the preferred PSSM data format akin to the one produced by PSI-BLAST, which is the format used by most published predictive models in the field. Therefore, in this paper, I present an optimized and fast hybrid alignment solution combining these two state-of-the-art methods to quickly generate extensive protein sequence alignments and their corresponding PSSM profiles. Sections are as follows: I explain the pipeline architecture components, give a runtime comparison between both tools (MMseqs2 vs PSI-BLAST), and finally, present the evaluation results of the EPSAPG output with regard to three different ML models (TMSEG [24], REPROF [9], and LocTree2 [25]) from Predict-Protein [26].

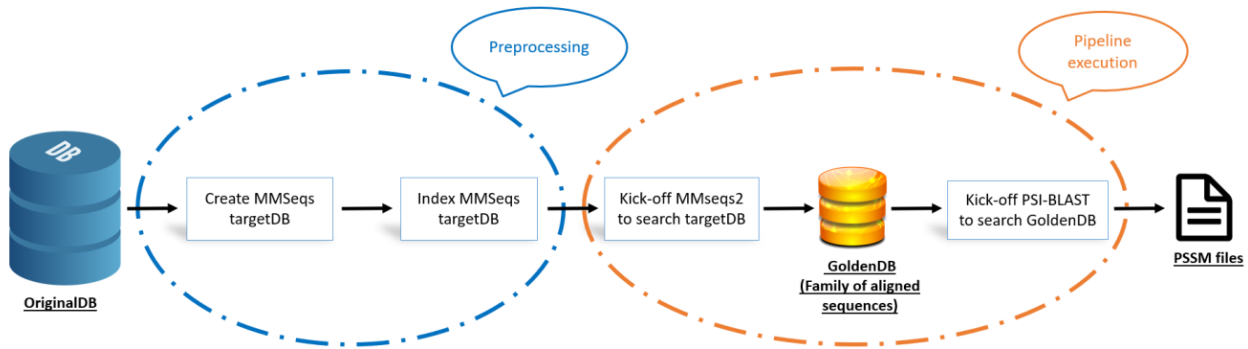

**Figure 1.** EPSAPG high-level architecture. The high-level architecture consists of two main parts: a preprocessing step, blue section, and a pipeline execution step, orange section. The preprocessing procedure compiles an MMseqs database, named targetDB, from the FASTA file of the original database to be searched. From the targetDB, a large index table is constructed that later will be used by the pipeline for a fast query alignment. The preprocessing step is crucial for a faster and reusable index table in future runs.

## II. EPSAPG ARCHITECTURE

EPSAPG combines MMseqs2 and PSI-BLAST to quickly generate profiles for a set of query sequences using heuristics. It has the ability to align batch query sequences in parallel and to generate corresponding PSI-BLAST PSSM profiles. To achieve this goal, preprocessing of the database to be searched is crucial before running EPSAPG (Figure 1). The preprocessing stage yields an index table, which may have a substantial size reaching hundreds of gigabytes, particularly when dealing with large-scale data sets. This step is essential to make the alignment faster afterward. It is important to underscore that the creation of an index table is not obligatory for querying a database using EPSAPG. The search command will automatically generate the index table at the commencement of the pipeline execution. Nonetheless, it is strongly advised to proactively generate and retain the index table in advance. By doing so, the search command will not need to construct a new index table for each query, particularly when multiple pipeline runs are being conducted on the same database.

The output of the preprocessing step (Figure 1 - blue) is an index table of the original target database to be searched. EPSAPG consumes then this index table and runs a set of consecutive modules in order to finally generate a PSSM profile (Figure 1 - orange). The first module to kick off is the "pre-filter module" (Figure 2). This module constructs the main component of MMseqs2 power. The pre-filtering module computes an ungapped alignment score for all consecutive *k-mer* matches between all query sequences and all database sequences and returns the highest score per sequence [2, 7, 23]. The prefilter *k-mer* match stage is key to high speed and sensitivity. It detects consecutive short words (i.e. *k-mer*) matching on the same diagonal. The diagonal of a *k-mer* match is the difference between the positions of the two similar *k-mer* in the query and in the target sequence [2, 7, 23].

Once the pre-filtering module is complete, it generates the best matches for each query sequence. Those matches do not go beyond the number N, the maximum number of sequences to be retrieved by the module. This value can be provided as a parameter to the pipeline, and it is set in the current presented study to 1000 aligned sequences per query sequence. The hits are then passed to the "alignment module" (Figure 2). This module implements a SIMD accelerated Smith-Waterman-alignment algorithm [27] of all sequences that pass a cut-off for the prefiltering score in the first module. It processes each sequence pair from the prefiltering results and aligns them in parallel, calculating one alignment per core at a single point in time. It processes each sequence pair from the prefiltering results and aligns them in parallel, calculating one alignment per core at a single point in time. This step is implemented using a vectorized programming algorithm that makes use of the SIMD instruction set. This is where the SSE4.1 instruction set requirement is essential. In the end, the module calculates alignment statistics such as sequence identity, alignment coverage, and e-value of the alignment [28], which are then passed to the "convertalis module" and the "parsimus module" to transform the intermediate results into FASTA format, using optimized algorithms inspired from and detailed in [29].

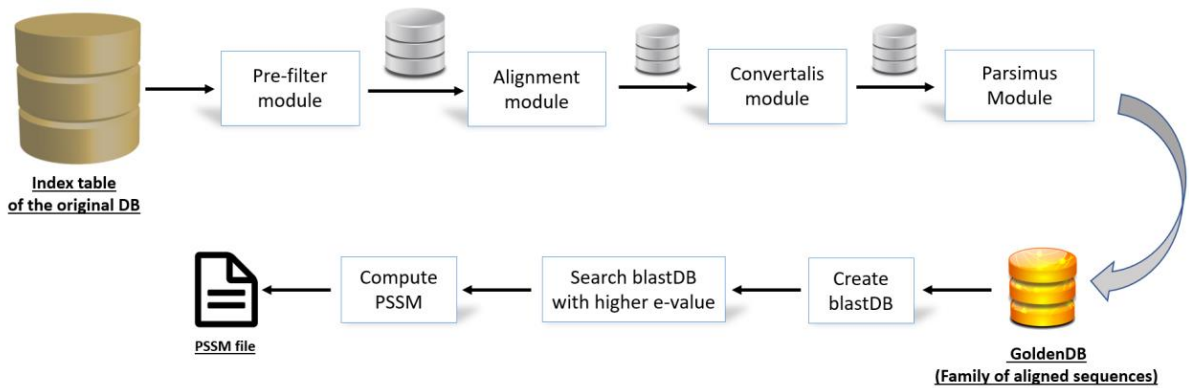

**Figure 2.** EPSAPG Low-Level Architecture. The Pipeline consumes an index table of the database to be searched as input then runs the (Pre-filter module) with a maximum number of sequences to be retrieved (a pipeline parameter - default 1000). The result is then aligned (Alignment module) to produce a result DB that will then be converted (Convertalis module) to an intermediate data set named epsapg.tuple. It is a large file containing a tab-separated list of 3 columns: query header, target header, and target sequence. The file is then parsed (Parsimus module) to create intermediate query and corresponding database files, in FASTA format. The query and database files will then be consumed to generate the final BLAST output (local alignment result, PSSM, or ASCII PSSM – a pipeline parameters to be provided) via PSI-BLAST search on the parsed golden database with a high e-value (a pipeline parameter set to the default value of 10)

I call this intermediate data set the golden DB, with a default maximum value of one thousand sequences per query sequence. This set of best homologous hits, for each query sequence, is passed to the subsequent PSI-BLAST modules, designed in the pipeline to run in parallel using the full power of the machine, to generate the final PSSM profiles.

### III. DATA

Runtime measurements in the results section refer to search alignments of amino acid sequences (excluding the time needed for indexing ~1hrs20min) against a sample of UniProt [30] Reference Cluster with 90% sequence identity (uniref90 2019_02). The pipeline has no limitations in terms of database size, but in order to assess the software's optimal performance, I needed an index table of a database that can fit entirely in the RAM of the test server. Since the whole uniref90 will generate an index table of roughly the size of 250 GB, a random sample, of ~69 million proteins from uniref90, was selected to perform the analysis. The sample generates a 181 GB index table, which can fit in the test machine memory.

All data, query sequences, search results, and additional supplementary online material (SOM) reported and referenced in this manuscript have been deposited to Zenodo at https://zenodo.org/record/8212007, and https://zenodo.org/record/8051133.

### IV. RESULTS AND DISCUSSION

All results presented in this section were conducted on a dedicated server granted by the University of Luxembourg. The machine's main characteristics are summarized as follows: a memory size of 198 GB with an Intel Xeon E312xx CPU of 4 cores supporting the SSE4.1 instruction set.

Within this section, I present an analysis on the impact of sequence length and query batch size on the runtime performance of each separate software, MMseqs2 and PSI-BLAST. Subsequently, I present an evaluation concerning the pipeline's output performance across three different protein ML predictive models. Finally, considerations regarding potential avenues for achieving enhanced execution speed of the pipeline will be discussed.

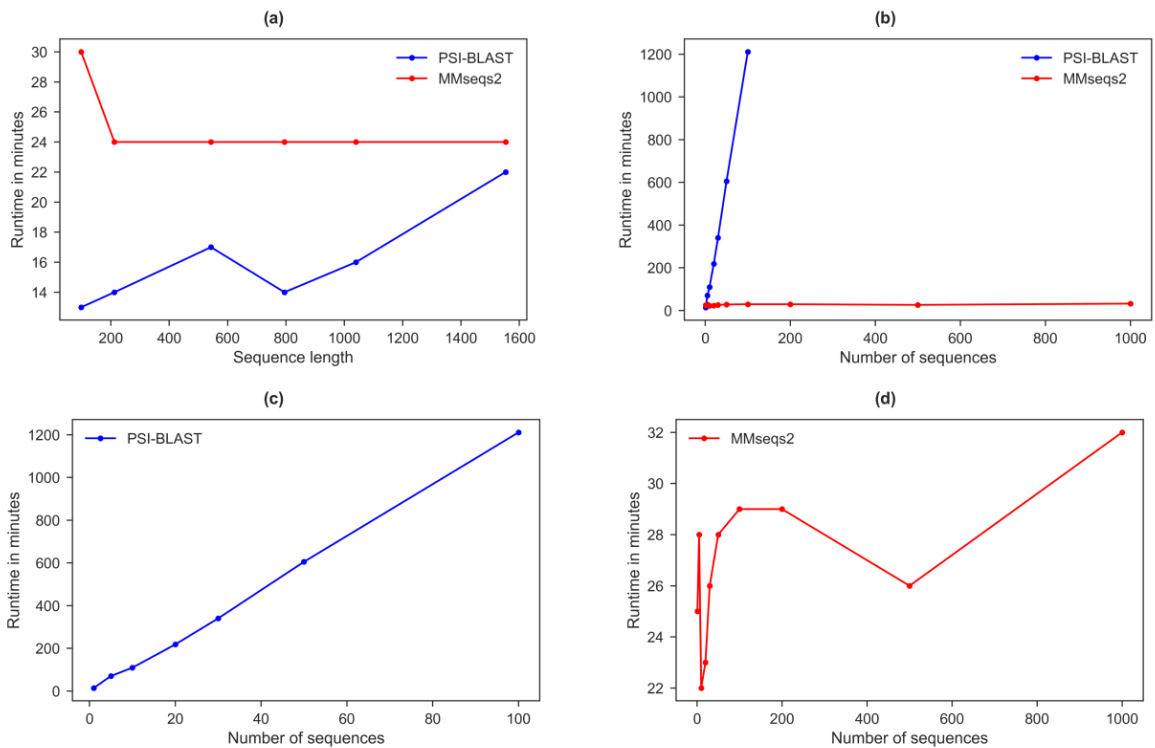

**Figure 3.** Runtime analysis results in a nutshell. (a) depicts the effect of sequence length on the runtime using each alignment method. The sequence length affects PSI-BLAST runtime while it does not for MMseqs2. (c) shows the effect of random bulk sequences run as a single query on the runtime of PSI-BLAST, while (d) the effect on the runtime of MMseqs2. (b) benchmarks, in a single figure, the impact of the number of sequences as a single query on the runtime of each alignment tool. It shows that PSI-BLAST scales linearly while MMseqs2 remains relatively constant.

### A. MMseqs2 VS PSI-BLAST: Runtime evaluation

Comparing the runtime execution of each method separately using a single query protein shows that the amino acid sequence length of the protein affects PSI-BLAST runtime while it does not for MMseqs2 (Figure 3-a). Another observation shows that PSI-BLAST runtime increases with the increasing number of residues in a query sequence, while for MMseqs2, the runtime remains constant (Figure 3-b). However, the experiments show the opposite for batch processing. The increasing number of sequences included in one query heavily affects PSI-BLAST runtime (see Figure 3-c). On the other hand, MMseqs2 is barely affected (see Figure 3-d). MMseqs2 runtime ranges between 22 and 32 minutes with an average runtime of ~30 minutes. PSI-BLAST runtime increases linearly, and this is mainly due to the sequential processing of the query sequences by the software.

PSI-BLAST is faster when it comes to a single query alignment while MMseqs2 scales much better in terms of batch sequence processing. This feature was exploited to build a faster batch sequence alignment pipeline to generate PSSM profiles.

### B. Pipeline Evaluation

One strategy to assess the performance of the pipeline involves evaluating the impact on predictions made by existing ML models using EPSAPG-generated profiles. This evaluation aims to determine whether the pipeline-generated profiles introduce any adverse effects on the predictions. The pipeline was tested on three ML methods, namely TMSEG [24], REPROF [9], and LocTree2 [25].

*1) TMSEG:* TMSEG [24] is a tool that predicts the transmembrane segments in a protein. For this method, 265 unique proteins were selected to be evaluated using both PSI-BLAST and EPSAPG. The initial metric dimension for evaluation is the runtime required to obtain the PSSM files. PSI-BLAST needed approximately 48.5 hours (refer to Table 1) to search the database and to generate the PSSMs, whereas EPSAPG accomplished this task within ~30 minutes, resulting in a significant reduction in runtime.

TABLE I. RUNTIME COMPARISON OF RUNNING A 265-QUERY BATCH AGAINST A RANDOM SAMPLE OF 68 MILLION PROTEINS FROM UNIREF90. PSI-BLAST NEEDS 48HRS 28MIN, ROUGHLY 100 TIMES MORE THAN EPSAPG 31MIN, TO GENERATE PSSM PROFILES.

| Number of sequences | EPSAPG runtime | PSI-BLAST runtime |
|---|---|---|
| 265 | 31 min 24 sec | 48 hrs 28 min 11 sec |

The second metric dimension to evaluate is the character-wise similarity between both outputs. Figure 4-b illustrates that the number of instances exhibiting a character-wise similarity percentage to the ground truth of 0.95 or higher, resulting from PSI-BLAST PSSMs, is lower compared to those resulting from EPSAPG, as depicted in Figure 3-a. Conversely, PSI-BLAST yields significantly higher results than EPSAPG when considering a similarity threshold of 0.9. From a statistical standpoint, when comparing the mean similarity values overall, EPSAPG results exhibit a mean similarity of 0.830 with a standard deviation of 0.125, while PSI-BLAST results demonstrate a mean similarity of 0.832 with a standard deviation of 0.119.

In general, results in Figure 4 show that both tools demonstrate comparable performance, with the distribution among all bars effectively compensating for discrepancies observed in specific regions in the compared tools. In terms of the predicted string using PSSMs from both methods, Figure 5 shows that most of the predicted pairs are the same with a similarity mean of 0.945 and a standard deviation of 0.061.

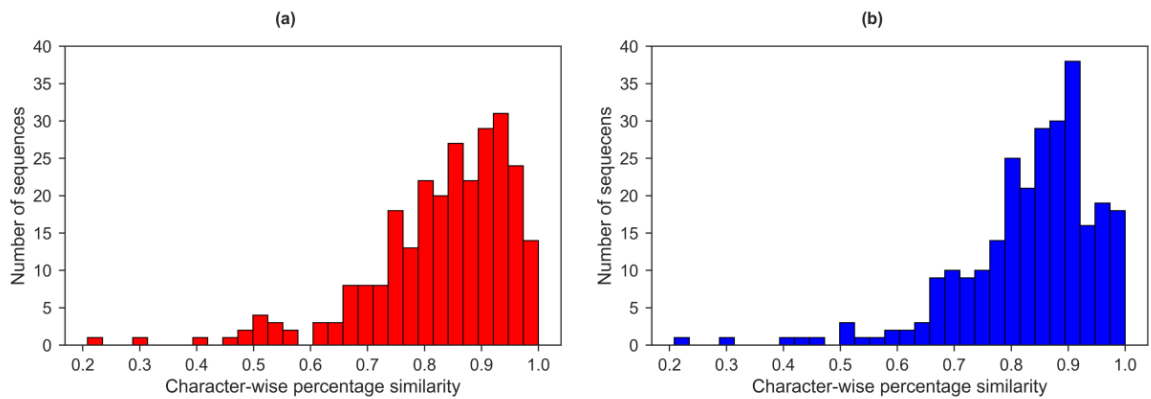

**Figure 4.** Character-wise TMSEG prediction similarity to the ground truth. (a) represents the percentage distribution of how similar the residue predictions of a sequence are to the ground truth labels using EPSAPG. (b) represents the percentage distribution of how similar the residue predictions of a sequence are to the ground truth labels using PSI-BLAST.

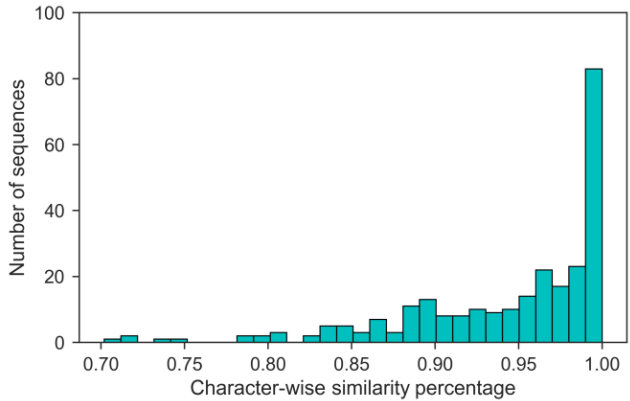

**Figure 5.** Character-wise similarity percentage between PSI-BLAST and EPSAPG PSSM based predicted transmembrane segments of the same amino acid sequence with respect to the ground truth.

To conduct further analyses of the output predictions derived from EPSAPG's profiles, two additional studies were carried out: a segment prediction analysis on the same test set and an additional analysis employing a larger data set consisting of 8817 proteins. The findings revealed a similar trend, wherein the runtime execution of EPSAPG was ~540 times faster compared to PSI-BLAST. (For further details consult section 1 of the SOM, as referenced in the data section)

2) *REPROF:* REPROF [9] is a tool that predicts a Q3 protein secondary structure. For this ML methodology, a set of 250 unique proteins was selected for evaluation using both PSI-BLAST and EPSAPG. The initial metric to be assessed is the runtime required to obtain the PSSM files. PSI-BLAST used up ~50 hours and a half (refer to Table 2) to search the target database and generate the profiles, while EPSAPG accomplished the same task within a mere half an hour, yielding a noteworthy reduction in runtime by approximately 100-fold.

TABLE II. RUNTIME COMPARISON OF RUNNING A 250-QUERY BATCH AGAINST A RANDOM SAMPLE OF 68 MILLION PROTEINS FROM UNIREF90. PSI-BLAST NEEDS 50HRS 33MIN, ROUGHLY 100 TIMES MORE THAN EPSAPG 32M, TO GENERATE PSSM PROFILES.

| Number of sequences | EPSAPG runtime | PSI-BLAST runtime |
|---|---|---|
| 250 | 32 min 33 sec | 50 hrs 33 min 57 sec |

Similar to TMSEG, a character-wise similarity analysis was conducted. Upon initial visual observation, the histograms generated by both alignment methods exhibit a distribution that can be deemed comparable. Furthermore, it is worth noting that the instances exhibiting a character-wise similarity percentage of 0.8 or higher to the ground truth, as a result of PSI-BLAST PSSMs (Figure 6-b), demonstrate a relatively lower magnitude compared to those resulting from EPSAPG (Figure 6-a). Overall, both histograms look comparable with some bars compensating for other differences. This observation is supported by the statistics computed to compare the overall similarity means. EPSAPG has a similarity mean of 0.781 with a standard deviation of 0.092, whereas PSI-BLAST has a similarity mean of 0.779 with a standard deviation of 0.092. Therefore, both tools have roughly similar performance.

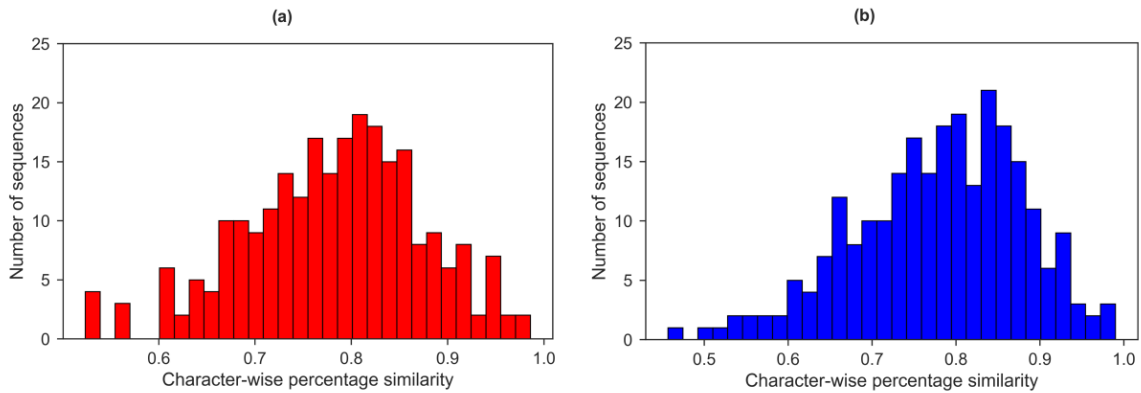

**Figure 6.** Character-wise REPROF prediction similarity to the ground truth. Histogram (a) represents the percentage distribution of how similar the residue predictions of a sequence are to the ground truth labels using EPSAPG. Histogram (b) represents the percentage distribution of how similar the residue predictions of a sequence are to the ground truth labels using PSI-BLAST.

Regarding the predicted secondary structure derived from PSSMs generated by either EPSAPG or PSI-BLAST aligned sequences families, Figure 7 demonstrates a substantial overlap in the majority of predicted pairs, exhibiting a similarity absolute difference mean of 0.0005 and an SD of 0.281. A larger data set, of 4692 proteins, analysis was also performed and showed a similar trend, with a runtime improvement of 414 times. (For further details consult section 2 of the SOM, as referenced in the data section).

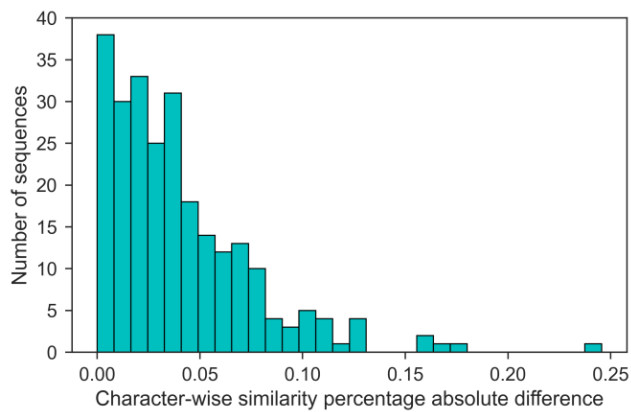

**Figure 7.** The absolute difference of the character-wise similarity percentage between PSI-BLAST and EPSAPG PSSM-based predicted secondary structure of the same amino acid sequence with respect to the ground truth.

3) *LocTree2:* LocTree2 [25] is a computational method utilized to predict the subcellular localization of proteins. This method employs an innovative approach that combines both sequence-based and homology-based strategies. It makes predictions for three target organisms: archaea, bacteria, and eukaryota. The classes that are predicted depend on the organism: 18 localization classes are predicted for eukaryota, 6 for bacteria, and 3 for archaea. For a thorough assessment of the effect of the pipeline's generated profiles on the predictions of LocTree2, I run 2 types of experiments on the organism with the largest number of classes, eukaryota. The first experiment made use of a balanced data set of 177 proteins, while the second experiment made use of a larger and random set of 600 proteins. The large test set was constrained by the prediction method runtime, which requires ~1.3 minutes per sequence, yielding a total prediction runtime of 13 hours for the 600 proteins. The test sets were extracted from eukaryota DeepLoc data set [31].

For the balanced data set, I generated profiles using both PSI-BLAST, which took ~38 hours, and EPSAPG, which took ~30 minutes. The confusion matrix results of LocTree2 for each input show that profiles generated from both tools exhibit a similar performance, with EPSAPG pipeline output providing slightly better results in the 3 best-predicted classes: 'nucleus-s', 'extracellular-s', and 'mitochondrion-s' (refer to figure S5 and figure S6 in the SOM as referenced in the data section). In terms of comprehensive accuracy, the utilization of EPSAPG yielded a score of 32.7%, whereas relying solely on PSI-BLAST resulted in an accuracy of 31.1%.

In the case of the expanded test set comprising 600 proteins, the evaluation was restricted to the pipeline alone. This decision was prompted by the fact that utilizing only PSI-BLAST would necessitate approximately five days to create the PSSM profiles. The outcomes reveal that the metrics presented for the smaller test set hold when compared to a larger data set, with the most accurately predicted three classes remaining the same: 'nucleus-s', 'extracellular-s', and 'mitochondrion-s'. Furthermore, the achieved overall accuracy stands at 35%, which is still comparable to the results of the smallest test set. A detailed graphical representation of these findings can be found in figure S7 of the SOM.

*C. Can we achieve an even faster sequence alignment?*

The runtime analysis of each module separately within the EPSAPG search framework revealed that the loading (I/O) of the index table constitutes approximately 90% of the overall reported runtime. This loading process is primarily constrained during the pre-filtering stage and the execution of the Convertalis module. Therefore, an approach that involves preloading the entire database into memory (RAM), prior to querying, and using it directly during the search and conversion of results, would bring about a significant reduction in runtime.

An effective strategy involves loading the whole database into the RAM and lock the pages within the memory space. Later in the search, I can map data from the locked DB instead of loading it again from the disk. The "vmtouch" command was used to serve this purpose. It is a utility used to manage and control the file system cache of Unix and Unix-like systems [32]. The tool loads the index table into the RAM and locks the pages there. This can be achieved through the following command:

*sudo /usr/local/bin/vmtouch -t -l -d targetDB.idx*

The command takes three arguments: "-t" to touch or load the file into the system cash, "-l" to lock the pages, and "-d" to run it as a daemon. Also, the command must have root privileges. The reason behind that is the "max lock memory" constraint enforced by the operating system, which is usually set to 64 kilobytes. Otherwise, the database will be loaded into the RAM but not locked there. This can be resolved by either including the full PATH into the sudo environment variables and running the process with root privileges or just running the command with sudo but using the full path to the executable. With this workaround, I was able to generate the PSSM file for an average length protein sequence (~450 amino acids) in 53 secs using EPSAPG.

## V. CONCLUSION

The findings of this research reveal that MMseqs2 exhibits slower execution time in the context of individual query searches as compared to PSI-BLAST. However, it demonstrates considerable efficiency in batch-processing scenarios. Leveraging this feature, I have combined MMseqs2 and PSI-BLAST modules to construct a pipeline named EPSAPG. This pipeline accelerates the search for sequence alignments, relying on MMseqs2 power, and computes corresponding PSSM profiles, which serve as vital input components in numerous ML predictive models. Notably, when handling substantial query batches of protein sequences, EPSAPG surpasses PSI-BLAST in terms of performance, achieving a speed improvement exceeding two orders of magnitude. I conducted a thorough assessment of the output results generated by the pipeline and compared them with those obtained from PSI-BLAST. The observations indicate that the performance of EPSAPG's output is at least comparable to, and in certain cases slightly superior to, the results obtained by PSI-BLAST. Subsequently, I introduced a workaround to achieve near-instant alignment for individual queries. Given enough computing resources and utilizing the "vmtouch" command, one can load the index table into the RAM and lock its pages, allowing much faster access through sequence mapping for single query searches. Ultimately, with EPSAPG I succeeded in achieving a significantly improved runtime speed of 53 seconds for an average sequence length of ~450 amino acids. The resulting alignments generate a PSSM profile that encompasses nearly identical statistical information to the matrix produced by PSI-BLAST, which typically takes tens of minutes for each query sequence relying only on PSI-BLAST.

Concerning future work, a beneficial added value would be the utilization of EPSAPG to construct a protein evolutionary information database. This database would encompass precomputed PSSM profiles for a large repository of distinct protein sequences, extensively employed by the scientific community. An example of such a cluster would be the UniRef30 [30] data set, which comprises tens of millions of unique proteins with less than 30% sequence similarity. This will also offer further stress tests assessing the pipeline's capabilities and potential to handle larger query sets. Furthermore, this database would serve as a robust foundation for the ongoing advancements in protein sequence large language models [33], with a focus on encoding the evolutionary information of protein sequences into their embeddings.

## ACKNOWLEDGMENTS

The author would like to thank Michael Bernhofer, Martin Steinegger, and Burkhard Rost for their insights and feedback. The author would like to also thank the University of Luxembourg for the computing resources granted for this project.